\documentclass[preprint,aps,showpacs,superscriptaddress,floatfix]{revtex4}
\usepackage{graphicx}
\newcommand{\eps}{\epsilon}
\newcommand{\om}{\omega}
\newcommand{\si}{\sigma}
\newcommand{\tauv}{\mbox{\boldmath$\tau$}}
\newcommand{\ra}{\rightarrow}
\newcommand{\beq}{\begin{equation}}
\newcommand{\eeq}{\end{equation}}
\newcommand{\beqa}{\begin{eqnarray}}
\newcommand{\eeqa}{\end{eqnarray}}
\newcommand{\bea}{\begin{array}}
\newcommand{\eea}{\end{array}}
\newcommand{\bite}{\begin{itemize}}
\newcommand{\eite}{\end{itemize}}
\newcommand{\bfig}{\begin{figure}}
\newcommand{\efig}{\end{figure}}
\newcommand{\btable}{\begin{table}}
\newcommand{\btabul}{\begin{tabular}}
\newcommand{\etabul}{\end{tabular}}
\newcommand{\etable}{\end{table}}
\newcommand{\ben}{\begin{enumerate}}
\newcommand{\een}{\end{enumerate}}
\newcommand{\lan}{\langle}
\newcommand{\ran}{\rangle}

\begin{document}
\title{Theory for a Hanbury Brown Twiss experiment with a ballistically expanding cloud of cold atoms}
\author{J. Viana Gomes}
\affiliation{Laboratoire Charles Fabry de l'Institut d'Optique,
UMR 8501 du CNRS et Universit\'{e} Paris 11, 91403 Orsay Cedex,
France} \affiliation{Departamento de Fisica, Universidade do
Minho, Campus de Gualtar, 4710-057 Braga, Portugal}
\author{A. Perrin}
\affiliation{Laboratoire Charles Fabry de l'Institut d'Optique,
UMR 8501 du CNRS et Universit\'{e} Paris 11, 91403 Orsay Cedex,
France}
\author{M. Schellekens}
\affiliation{Laboratoire Charles Fabry de l'Institut d'Optique,
UMR 8501 du CNRS et Universit\'{e} Paris 11, 91403 Orsay Cedex,
France}
\author{D. Boiron}
\email{denis.boiron@institutoptique.fr}
\homepage{http://www.atomoptic.fr} \affiliation{Laboratoire
Charles Fabry de l'Institut d'Optique, UMR 8501 du CNRS et
Universit\'{e} Paris 11, 91403 Orsay Cedex, France}
\author{C. I. Westbrook}
\affiliation{Laboratoire Charles Fabry de l'Institut d'Optique,
UMR 8501 du CNRS et Universit\'{e} Paris 11, 91403 Orsay Cedex,
France}
\author{M. Belsley}
\affiliation{Departamento de Fisica, Universidade do Minho, Campus
de Gualtar, 4710-057 Braga, Portugal}

\begin{abstract}
We have studied one-body and two-body correlation functions in a
ballistically expanding, non-interacting atomic cloud in the
presence of gravity. We find that the correlation functions are
equivalent to those at thermal equilibrium in the trap with an
appropriate rescaling of the coordinates. We derive simple
expressions for the correlation lengths and give some physical
interpretations. Finally a simple model to take into account
finite detector resolution is discussed.
\end{abstract}
\pacs{03.75.Hh, 05.30.Jp}

\maketitle

Whether a source emits photons or massive particles, if it is to
be used in an interferometric experiment, an essential property is
its coherence. The study of coherence in optics has shown that
more than one kind of coherence can be defined \cite{coh optic}.
The most familiar type of coherence is known as first order
coherence and is related to the visibility of interference fringes
in an interferometer. It is proportional to the value of the
correlation function of the associated field. Second order
coherence is less intuitive and corresponds to the correlation
function of the intensity or squared modulus of the field. From a
particle point of view, second order coherence is a way of
quantifying density correlations and is related to the probability
of finding one particle at a certain location given that another
particle is present at some other location. Particle correlations
can arise simply from exchange symmetry effects and exist even
when there is no interaction between the particles. This fact was
clearly demonstrated in the celebrated Hanbury Brown Twiss
experiment which showed a second order correlation for photons
coming from widely separated points in a thermal source such as a
star \cite{hbt}.

Analogous correlations in massive particles have also been
studied, particularly in the field of nuclear physics
\cite{baym,boal,jacak,gluon,Iannuzzi}. Spatial correlations using
low energy electrons have also been studied \cite{Henny99,
Oliver99}. The advent of laser and evaporative cooling techniques
has also made it possible to look for correlations between neutral
atoms and recently a wide variety of different situations have
been studied \cite{Yasuda96, Hellweg03, Greiner05,
Bloch05,Esslinger05, Esteve05, Schellekens}. Correlation phenomena
are generally richer when using massive particles because they can
be either Bosons or Fermions, they often have a more complex
internal structure and a large range of possible interactions with
each other. In the field of ultra-cold atoms, the many theoretical
papers to date have included treatments of bosons in a simple
three dimensional harmonic trap \cite{glauber, castin}, a 1D
bosonic cloud in the Thomas Fermi regime and Tonks-Girardeau limit
\cite{theo1D_shlya,theo1D_mora,theo1D_caz}, the Mott-insulator or
superfluid phase for atoms trapped in optical lattices
\cite{theoReseau} and the  2D gas \cite{theo2D}.

Almost all these theoretical treatments have dealt with atomic
clouds at thermal equilibrium. On the other hand, all the
experiments so far except Ref.\cite{Esteve05} have measured
correlations in clouds released from a trap which expand under the
influence of gravity and possibly interatomic interactions. It is
generally not trivial to know how the correlation properties
evolve during expansion. Moreover, matter waves have different
dispersion characteristics than light. All this raises interesting
questions concerning the value of the correlation lengths during
the atomic cloud expansion. In particular we would like to know
how to use the results of Ref.\cite{glauber} to analyze the
experimental results of Ref.\cite{Schellekens}, a conceptually
simple experiment in which second order correlations were measured
in a freely expanding cloud of metastable helium atoms. The
correlation length was defined as the characteristic length of the
normalized second order correlation function. We will use the same
definition in this paper (see section \ref{def} for details).

To illustrate a more general question that comes up in thinking
about the coherence of de Broglie waves, consider a beam of
particles with mean velocity $v$ hitting a detector. Two obvious
length scales come immediately to mind, the de Broglie wavelength
$\hbar/(m\Delta v)$ associated with the velocity spread $\Delta v$
and the length associated with the inverse of the energy spread of
the source $\hbar v/m(\Delta v)^2$. These two scales are obviously
very different if $v$ is large compared to the velocity spread. In
this paper, we will show that in an experiment such as
\cite{Schellekens}, the correlation length corresponds to neither
of the above length scales, although they can be relevant in other
situations. We find that the correlation length after an expansion
time $t$ of a cloud of initial size $s$ is $\hbar t/ms$. This
result is the atom optical analog of the van Cittert-Zernike
theorem \cite{Born:1980}. It has also been stated in a different
form in Ref. \cite{Miller}. For the special case of an ideal gas
in a harmonic trap of oscillation frequency $\om$, the correlation
length can be recast as $\lambda\om t$ where $\lambda$ is the
thermal de Broglie wavelength. Hence the correlation length after
expansion is simply dilated compared to that at equilibrium with
the same scaling factor as the spatial extent of the cloud itself.

We will confine ourselves here to the case of a cloud of non
interacting atoms released suddenly from a harmonic trap. The
paper is organized as follows. We will begin in section \ref{sec
eq} with some simple definitions and  general results about the
correlation properties of a non-interacting cloud both at thermal
equilibrium in a trapping potential and after a ballistic
expansion. Without making any assumptions about the form of the
trapping potential, we can only find simple analytical results in
the limit of a non-degenerate gas. Next we will make a more exact
and careful treatment by specializing to the very important case
of a harmonic potential. We introduce the flux operator
\cite{Whitlock} involved in the experimental electronic detection
with metastable helium and then calculate the correlation function
of the flux. We will summarize the results and give a physical
interpretation in section \ref{sec interpret}. This interpretation
will allow us to comment on the rather different case of a
continuous beam as in the experiments of
Ref.\cite{Yasuda96,Esslinger05,Iannuzzi}. In section \ref {sec
exp} we will use our results to analyse the experimentally
important problem of finite detector resolution. Finally, the
appendix adds some detailed calculations concerning the
expressions found in section \ref{sec expansion}.

\section{General results on correlation functions of non-interacting gases}\label{sec eq}

Here we recall some basic results concerning the
density and first and second order correlation functions for a
cloud of non-interacting bosons at thermal equilibrium. A more
detailed analysis can be found in Ref.\cite{glauber}. Theoretical
treatments that take into account interatomic interactions can be
found in Ref.\cite{glauber, castin, krauth}. We also give some
approximate results for a non-interacting gas after it has
expanded from a trap.

\subsection{Definitions}\label{def}
Consider a cloud of $N$ atoms at thermal equilibrium at a
temperature $T$, confined in a trapping potential. This potential
is characterized by $\{\eps_{\bf j},\psi^0_{\bf j}({\bf r})\}$ the
energy and wavefunction of level ${\bf j}$ (here supposed
non-degenerate for simplicity).
In second quantization, one defines the field operators

$$
\hat{\Psi}^\dag({\bf r})=
 \sum_{\bf j}\psi_{\bf j}^*({\bf r})\hat{a}^\dag_{\bf j}~,~~~
\hat{\Psi}({\bf r})=
 \sum_{\bf j}\psi_{\bf j}({\bf r})\hat{a}_{\bf j}.
$$

\noindent The operator $\hat{a}^\dag_{\bf j}$ creates and $\hat{a}_{\bf j}$
annihilates one particle in state $|\psi_{\bf j}\ran$ whereas
$\hat{\Psi}^\dag({\bf r})$ creates and $\hat{\Psi}({\bf r})$
annihilates a particle at position $\bf r$.\\

Correlation functions and the atomic density are statistical
averages of such field operators. We use the Bose-Einstein
distribution, $\langle \hat{a}_{\bf j}^\dag \hat{a}_{\bf
k}\rangle=\delta_{\bf j\;k}(e^{\beta(\eps_{\bf j}-\mu)}-1)^{-1}$
where $\beta=1/(k_BT)$, $k_B$ is the Boltzmann constant and $\mu$
is the chemical potential. The value of $\mu$ ensures the
normalization $\sum\limits_{\bf j }\langle \hat{a}_{\bf j}^\dag
\hat{a}_{\bf j}\rangle=N$. We can then define

\bite

\item the first order correlation function
$ G^{(1)}({\bf r},{\bf r'})= \langle\hat{\Psi}^\dag({\bf r})
       \hat{\Psi}({\bf r'})\rangle$,

\item the second order correlation function
$ G^{(2)}({\bf r},{\bf r'})= \langle\hat{\Psi}^\dag({\bf
r})\hat{\Psi}({\bf r})\hat{\Psi}^\dag({\bf r'})
       \hat{\Psi}({\bf r'})
\rangle $

\item and the density $\rho_{eq}({\bf r})=\langle\hat{\Psi}^\dag({\bf r})
       \hat{\Psi}({\bf r})\rangle=G^{(1)}({\bf r},{\bf r})$.

\eite

Several other first and second order correlation functions can be
defined (see below) but these are the most common ones. The first
order correlation function appears in interference experiments
whereas second order correlation functions are related to
intensity interference or density fluctuation. First and
second-order correlation functions are connected for thermal
non-interacting atomic clouds. The $G^{(2)}$ function contains a
statistical average of the type $ \lan\hat{a}_{\bf j}^\dag
\hat{a}_{\bf k}\hat{a}^\dag_{\bf l} \hat{a}_{\bf n}\ran$ which can
be calculated through the thermal averaging procedure (Wick
theorem \cite{wick}). One finds $ \lan\hat{a}_{\bf j}^\dag
\hat{a}_{\bf k}\hat{a}^\dag_{\bf l} \hat{a}_{\bf n}\ran = \lan
\hat{a}_{\bf j}^\dag \hat{a}_{\bf j}\ran\lan\hat{a}_{\bf k}^\dag
\hat{a}_{\bf k}\ran(\delta_{\bf jl}\delta_{\bf kn} +\delta_{\bf
jn}\delta_{\bf kl})+\lan \hat{a}_{\bf j}^\dag \hat{a}_{\bf j}\ran
\delta_{\bf kl}\delta_{\bf jn} $, which leads to $$ G^{(2)}({\bf
r},{\bf r'})=\rho_{eq}({\bf r})\rho_{eq}({\bf r'})+|G^{(1)}({\bf
r},{\bf r'})|^2 +\rho_{eq}({\bf r})\delta({\bf r-r'})$$ The last
term is the so-called shot-noise term. It will be neglected in the
following because it is proportional to $N$ whereas the others are
proportional to $N^2$.

It is convenient to define a normalized second order correlation
function

$$ g^{(2)}({\bf r},{\bf r'})={G^{(2)}({\bf
r},{\bf r'})\over \rho_{eq}({\bf r})\rho_{eq}({\bf r'})}.$$

\noindent If the cloud has a finite correlation length, then for
distances larger than this length the first-order correlation
function vanishes. Then $g^{(2)}({\bf r},{\bf r})=2$ and
$g^{(2)}({\bf r},{\bf r'})\ra 1$ when $|{\bf r-r' }|\ra\infty$.
This means that the probability of finding two particles close to
each other is enhanced by a factor of 2, compared to the situation
where they are far apart. This is the famous bunching effect first
observed by Hanbury Brown and Twiss with light \cite{hbt}.

The above expression of the $G^{(2)}$ function cannot be applied
in the vicinity and below the Bose-Einstein transition
temperature. The calculation of $ \lan\hat{a}_{\bf j}^\dag
\hat{a}_{\bf k}\hat{a}^\dag_{\bf l} \hat{a}_{\bf n}\ran$ is
performed in the grand canonical ensemble which assumes the
existence of a particle reservoir that does not exist for the
condensate. It is well known \cite{landau} that this gives
unphysically large fluctuations of the condensate at low enough
temperature. This pathology disappears at the thermodynamic limit
if there is an interatomic interaction \cite{landau}. It has also
been shown that it cancels for a finite number of non-interacting
particles if one uses the more realistic canonical ensemble
\cite{Politzer96}. One way to keep using the grand canonical
ensemble is to add the canonical result for the ground-state
\cite{glauber}. This approach is validated by the results in
Ref.\cite{Politzer96} and will be used in the following. The
largest deviation is expected to occur near the transition
temperature \cite{Politzer96}. The contribution of the ground
state is $-\lan \hat{a}_{\bf 0}^\dag \hat{a}_{\bf
0}\ran^2\delta_{\bf j0}\delta_{\bf k0}\delta_{\bf l0}\delta_{\bf
n0}$. Then, with $\rho_0$ the ground-state density, it follows
that,

\beq G^{(2)}({\bf r},{\bf r'})=\rho_{eq}({\bf r})\rho_{eq}({\bf
r'})+|G^{(1)}({\bf r},{\bf r'})|^2-\rho_0({\bf r})\rho_0({\bf
r'})\label{expr-g2} \eeq

The normalized second order then becomes $$g^{(2)}({\bf r},{\bf
r'})=1+\frac{|G^{(1)}({\bf r},{\bf r'})|^2}{\rho_{eq}({\bf
r})\rho_{eq}({\bf r'})}-\frac{\rho_0({\bf r})\rho_0({\bf
r'})}{\rho_{eq}({\bf r})\rho_{eq}({\bf r'})}$$

Because the ground state density is negligible for a thermal
cloud, the normalized correlation function $g^{(2)}({\bf r},{\bf
r'})$ still goes from 2 to 1 as the separation of ${\bf r}$ and
${\bf r'}$ increases. On the other hand, for a BEC at $T=0$, only
the ground-state is occupied. Then $|G^{(1)}({\bf r},{\bf
r'})|^2=\rho_{eq}({\bf r})\rho_{eq}({\bf r'})=\rho_0({\bf
r})\rho_0({\bf r'})$ and $g^{(2)}({\bf r},{\bf r'})=1$. The amount
of particle bunching present in the second order correlation
function can be quantified as $g^{(2)}({\bf r},{\bf r'})-1$ and
this typically decays exponentially as the modulus squared of the
separation between the two points increases. We define the
correlation length to be the characteristic length over which the
amount of particle bunching decays, that is the distance over
which $g^{(2)}({\bf r},{\bf r'})-1$ decays to $1/e$ of its maximum
value. The correlation length of a BEC is infinite. Such a system
is said to exhibit bunching at high temperature over the
correlation length and no bunching in the condensed phase.

\subsection{Correlations in an expanding cloud}\label{corr momentum}

In most experiments, particle correlations and other
characteristics  are not directly measured in the atom cloud,
(Ref. \cite{Esteve05} is an exception). Rather, the cloud is
released from a trap and allowed to expand during a ``time of
flight" before detection. For a sufficiently long time of flight,
and neglecting interactions between the atoms, the positions one
measures at a detector reflect the initial momenta of the
particles. The results of section \ref{def} concerning the
correlation functions in position space all have analogs in
momentum space. In fact the correlation functions in the two
reciprocal spaces are closely related. At equilibrium, i.e. inside
the trap, the following relationships can be easily derived:
$$\int d{\bf p}\;G^{(1)}({\bf p,p})e^{-i{\bf p.r}/\hbar}=\int d{\bf
R}\;G^{(1)}({\bf R-r/2,R+r/2})$$
$$\int d{\bf r}\;G^{(1)}({\bf r,r})e^{i{\bf q.r}/\hbar}=\int d{\bf
P}\;G^{(1)}({\bf P-q/2,P+q/2})$$

In other words, the spatial correlation length is related to the
width of the momentum distribution and the momentum correlation
length is related to the width of the spatial distribution {\it
i.e.} the size of the cloud. No equally simple and general
relationship holds for the second order correlation functions.
This is because, close to the BEC transition temperature, and at
points where the ground state wave function is not negligible, the
special contribution of the ground state, the last term in Eq.
\ref{expr-g2} must be included, and this contribution depends on
the details of the confining potential. On the other hand, for an
ideal gas far from the transition temperature one can neglect the
ground state density, make the approximation that the correlation
length is very short, neglect commutators such as $[{\bf \hat
r,\hat p}]$, and then write the thermal density operator as
$\hat\sigma = e^{-\beta{{\bf \hat P ^2}\over 2m}} e^{-\beta
V({\bf\hat r})}$. These approximations lead to:
$$
    G^{(2)}({\bf p},{\bf p'})=\rho_{eq}({\bf p})\rho_{eq}({\bf
    p'})+|G^{(1)}({\bf p},{\bf p'})|^2
$$
and,
$$
    G^{(1)}({\bf P-q/2,P+q/2})\sim
    e^{-\beta{{\bf P ^2}\over 2m}}\int d{\bf r}\;e^{-\beta V({\bf
r})}e^{i{{\bf q.r}\over\hbar}}
$$

One sees that in this limit, the interesting part of $G^{(2)}$ in
momentum space is proportional to the square of the Fourier
transform of the density distribution and independent of the mean
momentum ${\bf P}$. This result is the analog of the van
Cittert-Zernike theorem \cite{Born:1980}. For a trapped cloud of
size $s_\alpha$ in the $\alpha$ direction, one has a momentum
correlation ``length" given by: \beq p_\alpha^{(coh)}={\hbar\over
s_\alpha}. \eeq If atoms are suddenly released from a trap and
allowed to freely evolve for a sufficiently long time $t$, the
positions of the particles reflect their initial momenta and the
spatial correlation length at a detector is given by \beq
l_\alpha^{(d)}={p_\alpha^{(coh)} \over m} t ={\hbar t \over m
s_\alpha} \label{p coh length} \eeq

The normalized second order correlation function is then a
Gaussian of rms width $l^{(d)}/\sqrt 2$. This result was
experimentally confirmed in Ref. \cite{Schellekens}. One wonders
however, to what extent the approximations we have made are valid.
The clouds used in Ref. \cite{Schellekens} were in fact very close
to the transition temperature so that effects due to the Bose
nature of the density matrix may be important. Although the time
of flight was very long, it is useful to quantify the extent to
which identifying the momentum correlation length in the trap with
the spatial correlation length at the detector is accurate.
Finally, the effect of gravity on the falling atoms never appears
in the above approximate treatment, and we would like to clarify
the role it plays. In order to answer these questions we undertake
a more careful calculation. We will confine ourselves to atoms
initially confined in a harmonic trap, a good approximation to the
potential used in most experiments, and happily, one for which the
eigenstates and energies are known exactly.

\section{Density and correlation functions for a harmonic trap}\label{harm trap}

\subsection{At equilibrium in the trap}

The eigenfunctions for a 3-dimensional harmonic potential of oscillation frequency $\om_\alpha$
in the $\alpha$ direction, are given by:
$$ \psi^0_{\bf j}({\bf r})
            =\prod\limits_{\alpha=x,y,z}
            A_{j_\alpha}~e^{-{r_{\alpha}^2\over 2\sigma_\alpha^2}}
            ~H_{j_\alpha}(r_{\alpha}/\sigma_\alpha).$$

\noindent Here $\si_\alpha=\sqrt{\hbar\over m\om_\alpha}$ is the
harmonic oscillator ground-state size, $H_{j_\alpha}$ is the
Hermite polynomial of order $j_\alpha$ and
$A_{j_\alpha}=(\sqrt{\pi}\si_\alpha
2^{j_\alpha}(j_\alpha)!)^{-1/2}$. The eigenenergies are given by
$\eps_{\bf
j}=\sum\limits_{\alpha=x,y,z}\hbar\om_\alpha(j_\alpha+1/2)$. Then
\cite{landau,glauber}, with $\tau_\alpha=\beta\hbar\om_\alpha$ and
$\tilde\mu=\mu-\hbar\sum\om_\alpha/2$, one finds:
$$\rho_{eq}({\bf r})={1\over\pi^{3/2}}\sum\limits_{l=1}^\infty
e^{\beta
l\tilde\mu}\prod\limits_\alpha{1\over\si_\alpha\sqrt{1-e^{-2\tau_\alpha
l}}}\,e^{-\tanh({\tau_\alpha l\over 2}
){r_\alpha^2\over\si^2_\alpha}}
$$
and
$$G^{(1)}({\bf r},{\bf r'})={1\over\pi^{3/2}}\sum\limits_{l=1}^\infty
e^{\beta
l\tilde\mu}\prod\limits_\alpha{1\over\si_\alpha\sqrt{1-e^{-2\tau_\alpha
l}}}\,\exp\left[-\tanh({\tau_\alpha l\over 2}
)\left({r_\alpha+r'_\alpha\over 2
\si_\alpha}\right)^2-\coth({\tau_\alpha l\over 2}
)\left({r_\alpha-r'_\alpha\over 2 \si_\alpha}\right)^2\right].
$$

The above expressions can be transformed into more familiar forms
in limiting cases:

\bite

\item For high temperature, $\mu\ra -\infty$ and one recovers the Maxwell-Boltzmann
distribution. The density is $\rho_{eq}({\bf
r})={N\over\lambda^3}\prod\limits_\alpha \tau_\alpha
e^{-{\tau_\alpha \over 2} {r_\alpha^2\over\si_\alpha^2}}$ with
$\lambda={\hbar\sqrt{2\pi}\over\sqrt{mk_BT}}$ the thermal de
Broglie wavelength. The size of the cloud is
$s_\alpha=\si_\alpha/\sqrt{\tau_\alpha}=\sqrt{k_BT\over
m\om_\alpha^2}$.

The first order correlation function is \beq G^{(1)}({\bf r},{\bf
r'})={N\over\lambda^3}\prod\limits_\alpha \tau_\alpha
e^{-{\tau_\alpha \over 2} ({{\bf r_\alpha}+{\bf r'_\alpha}\over
2\si_\alpha})^2}e^{-\pi({{\bf r_\alpha}-{\bf r'_\alpha}\over
\lambda})^2}. \eeq\label{g1nondegen} Using our definition, the
correlation length is $l^{(t)}=\lambda/\sqrt{2\pi}$.

\item For a temperature close to but above the Bose-Einstein transition temperature,
one has to keep the summation over the index $l$. The density is
$\rho_{eq}({\bf r})={1\over\lambda^3}g_{3/2}[e^{\beta \tilde\mu}
\prod\limits_\alpha e^{-{\tau_\alpha \over 2}
{r_\alpha^2\over\si_\alpha^2}}]$ where
$g_a(x)=\sum\limits_{l=1}^\infty x^l/l^a$ is a Bose function. The
first order correlation function is $$G^{(1)}({\bf r},{\bf
r'})={1\over\lambda^3}\sum\limits_{l=1}^\infty {e^{l\beta
\tilde\mu}\over l^{3/2}}\prod\limits_\alpha e^{-{\tau_\alpha
l\over 2}({{\bf r_\alpha}+{\bf r'_\alpha}\over
2\si_\alpha})^2}e^{-{\pi\over l}({{\bf r_\alpha}-{\bf
r'_\alpha}\over \lambda})^2}.$$

As the temperature decreases, the number of
values of $l$ that contribute significantly to the sum increases. It is
then clear from the above expression for $G^{(1)}$ that the
correlation length near the center of the trap will increase and
that the normalized correlation function is no longer Gaussian.
Far from the center,  only the $l=1$ term is important and
the correlation function remains Gaussian. Thus close to degeneracy
the correlation length is position-dependent (for an explicit example see Sec.\ref{2nd_order}).

\item Near and below the transition temperature, the second order correlation
function is given by Eq. (\ref{expr-g2}) with $\rho_0({\bf
r})={e^{\beta \tilde\mu}\over 1-e^{\beta
\tilde\mu}}\prod\limits_\alpha{e^{-r_\alpha^2/\si_\alpha^2}\over
(\sqrt{\pi}\si_\alpha)^3 }$. As the temperature decreases, the
correlation at zero distance, $g^{(2)}(0,0)$ decreases from 2 to 1
and the correlation length increases. Around the transition
temperature, $g^{(2)}(0,0)$ is already significantly different
from 2 since the condensate peak density is already very large for
a non-interacting harmonically trapped cloud \cite{Hoppeler}. At
$T=0$, the correlation length is infinite and $g^{(2)}({\bf
r,r'})=1$.

\eite

\subsection{Correlations in a harmonically trapped cloud after expansion}\label{sec expansion}

Here we consider the cloud after expansion. First we discuss two
classes of detection methods which must be distinguished before
calculating correlation functions.

\subsubsection{Detection}

We assume that the trapping potential is switched off instantaneously at $t=0$.
The cloud expands and falls due to gravity. Two types of detection
can be performed: \bite

\item {\it Snap shot}. An image is taken of the entire cloud at $t=t_0$. We have then access to

$ G_{im.}^{(2)}({\bf r},t_0;{\bf r'},t_0)=
\langle\hat{\Psi}^\dag({\bf r},t_0)\hat{\Psi}({\bf r},t_0)
       \hat{\Psi}^\dag({\bf r'},t_0)\hat{\Psi}({\bf r'},t_0)
\rangle$

The usual imaging technique is absorption, and so one has access
to the above correlation functions integrated along the imaging
beam axis. This was used for the experiments of
Ref.\cite{Bloch05,Greiner05}.

\item {\it Flux measurement}. The atoms are detected when they cross a given plane. We will only consider the situation
in which this plane is horizontal at $z=H$. One has access to

$ G_{fl.}^{(2)}({\bf r}=\{x,y,z=H\},t;{\bf
r'}=\{x',y',z'=H\},t')=\lan \hat I({\bf r},t)\hat I({\bf
r'},t')\ran $

where $\hat I$ is the flux operator defined below. The detection
systems required for such experiments correspond most closely to
those of Refs. \cite{Yasuda96,Schellekens}, in which a
micro-channel plate, situated below the trapped cloud, recorded
the arrival times and in one case the positions of the atoms. It
also corresponds closely to imaging a cloud that crosses a thin
sheet of light \cite{Kasevich}, or to the experiment of
Ref.\cite{Esslinger05}, in which the transmission of a high
finesse optical cavity records atoms as they cross the beam. \eite

\noindent These two correlation functions are different, but if
the detection is performed after a long time of flight, they are in fact nearly equivalent. This
equivalence will be discussed in the following.\\

The flux operator is defined quantum-mechanically by
$$ \hat I({\bf r},t)={\hbar\over
m}\textit{Im}\left[\hat{\Psi}^\dag({\bf r},t)
               \partial_z\hat{\Psi}({\bf r},t)\right]={\hbar\over
2mi}\left[\hat{\Psi}^\dag({\bf r},t)\partial_z\hat{\Psi}({\bf
r},t)-\partial_z\hat{\Psi}^\dag({\bf r},t)\hat{\Psi}({\bf
r},t)\right] $$

\noindent The flux has thus the dimensions of a density times a velocity. We
will give the explicit expression of this velocity in the section
\ref{mean flux}. Here, the atomic field operators $\hat{\Psi}({\bf
r},t)$ depend on space coordinates as well as on time. They
represent the time evolution of the atomic field during the flight
of the atoms, falling from the trap. The field operators for the
falling cloud can be easily derived if we assume that there are no
interactions between the atoms and that the occupation number in
each mode is constant (as in free expansion). In this case, these
operators can be defined as
$$
\hat{\Psi}^\dag({\bf r},t)=
 \sum_{\bf j}\psi_{\bf j}^*({\bf r},t)\hat{a}^\dag_{\bf j}~,~~~
\hat{\Psi}({\bf r},t)=
 \sum_{\bf j}\psi_{\bf j}({\bf r},t)\hat{a}_{\bf j}
$$
where the spatiotemporal dependence is carried by the wave
function and the statistical occupation by the creation and
annihilation operators.

\subsubsection{Ballistic expansion of a harmonic oscillator stationary state}

After switching off the trap, the harmonic oscillator
wave-functions noted $\psi_{\bf j}^0$ are no longer stationary
states. There are two ways to calculate the correlation after
expansion: propagation of wavefunctions or propagation of the
density matrix (the Schr\"{o}dinger or the Heisenberg picture). In the
following we will use the first approach which is physically more
transparent (see \cite{cours_CCT} for the Heisenberg picture).

The ballistic expansion of a cloud is easy to calculate with the
appropriate Green function. The Green function $K$ is defined as
$$
\psi_{\bf j}({\bf r},t)=
  \int_{-\infty}^\infty {\bf dr}_0~K({\bf r},t;{\bf r}_0,t_0)~\psi^0_{\bf j}({\bf r}_0,t_0).
$$

\noindent As the $\psi^0_{\bf j}$ functions are stationary states for $t<0$,
we can take $t_0=0$ in the following. The Green function for
particles in an arbitrarily time-varying quadratic potential is
known \cite{fct_green}. After expansion, the  potential is only
due to gravity and the Green function is then

$$ K({\bf r},t;{\bf r}_0)= \left({m\over 2 i \pi\hbar t}\right)^{3/2} e^{i a({\bf r}-{\bf r}_0)^2}
e^{i b(z+z_0)}e^{-i c} $$ with $a={m\over 2\hbar t}, b={mgt\over
2\hbar}$ and $c={mg^2t^3\over 24\hbar}$.

One can then derive an analytical expression of $ \psi_{\bf
j}({\bf r},t)$ \cite{math,naraschewski}:

\beq \psi_{\bf j}({\bf r},t)= e^{i\phi({\bf r},t)}
\prod\limits_\alpha{e^{i
j_\alpha(\delta_\alpha+3\pi/2)}\over\sqrt{\om_\alpha t-i}}
\psi^0_{\bf j}({\bf \tilde r})\label{evol_temps}\eeq

\noindent where $\delta_\alpha=\tan^{-1}[{1\over\om_\alpha t}]$,
\beq\phi({\bf r},t)={m\over 2\hbar t}\left[(\tilde x \om_x
t)^2+(\tilde y \om_y t )^2+(\tilde z \om_z
t)^2+2gt^2(z-\frac{1}{8}gt^2) \right]-c-{3\pi\over 4} \label{phase
G1}\eeq

\noindent and, with ${\bf \tilde r}=\{\tilde x,\tilde y,\tilde z\}$, \beq\tilde
x={x\over \sqrt{1+\om_x^2t^2}},\tilde y={y\over
\sqrt{1+\om_y^2t^2}}, \tilde z={H-{1\over 2}gt^2\over
\sqrt{1+\om_z^2t^2}}\label{scaling}\eeq

In the case of flux measurement, the position of the detector is
fixed at $z=H$. The phase $\phi(\tilde x,\tilde y,t)$ is global as
it does not depend on the index $\bf j$; it will cancel in second
order correlation measurements. This is in contrast to
interferometric measurements where it is this phase that gives
rise to fringes. The above results show that after release, the
wavefunction is identical to that in the trap except for a phase
factor and a scaling factor in the positions \cite{castin_96}.
This scaling is obviously a property of a harmonic potential, and
it considerably simplifies the expression of the correlation
functions as we will see below.

\subsubsection{Flux operator}

Using $\partial_zH_n(z)=2nH_{n-1}(z)$, the spatial derivative of
the wavefunction can be written:

$$
\partial_z\psi_{\bf j}({\bf r},t)=
 {m\over\hbar}\Big\{[i v_2-v_1]\psi_{j_z}(z,t)-i v_3\sqrt{j_z}\, \psi_{j_z-1}(z,t)\Big\}
 \psi_{j_x}(x,t)\psi_{j_y}(y,t)$$

\noindent where the velocities $v_1$ $v_2$ and $v_3$ are time
dependent and are given by

\beqa
v_1(t)&=&\omega_z\frac{H-{1\over 2}gt^2}{1+\omega_z^2t^2}\\
v_2(t)&=&{1\over t}\left[H+{1\over 2}gt^2-\frac{H-{1\over
2}gt^2}{1+\omega_z^2t^2}\right]\\
 v_3(t)&=&\frac{\sqrt{2}\omega_z\sigma_z
}{\sqrt{1+\omega_z^2t^2}}\;e^{i\delta_z} \eeqa
The velocity $v_2$
is usually much larger than the other two and will give the
dominant contribution for the mean flux and the second order
correlation function. An atom with zero initial velocity will
acquire after a time $t$ a velocity $gt$ which is close to
$v_2(t)$. The flux operator is,
\beq
 \hat I({\bf r},t)=
    \sum\limits_ {\bf j,k}
       \left[v_2\psi_{\bf j}^*\psi_{\bf k}-{1\over 2}\left(v_3\sqrt{k}\,\psi_{\bf j}^*\psi_{\bf k-1_z}
       +v_3^*\sqrt{j}\,\psi_{\bf j-1_z}^*\psi_{\bf k}  \right)\right]
       ~\hat{a}_{\bf j}^\dag \hat{a}_{\bf k}\label{op_flux}\eeq
where ${\bf j-1_z}$ is the vector $(j_x,j_y,j_z-1)$ and where we
write $\psi=\psi({\bf r},t)$.

\subsubsection{Mean density and mean flux}\label{mean flux}

We will first calculate the mean density $\rho({\bf
r},t)=\lan\hat\Psi^\dag({\bf r},t)\hat\Psi({\bf r},t)\ran$. Using
Eq.(\ref{evol_temps}), one finds easily that $ \rho({\bf
r},t)={1\over\prod\limits_\alpha
\sqrt{1+\om_\alpha^2t^2}}\;\rho_{eq}({\bf\tilde r})$. This means
that the density has the same form during expansion up to an
anisotropic scale factor given by Eq.(\ref{scaling})
\cite{castin_96,shlya_96}. The statistical average of
Eq.(\ref{op_flux}) leads to
$$ \langle \hat I({\bf r},t)\rangle=
    \sum\limits_ {\bf j}
       \left[v_2|\psi_{\bf j}|^2-{\sqrt{j_z}\over 2}\left(v_3\psi_{j_z}^*\psi_{j_z-1}
       +v_3^*\psi_{j_z}\psi_{j_z-1}^*\right)
       |\psi_{j_x}\psi_{j_y}|^2\right]~\langle
        \hat{a}_{\bf j}^\dag \hat{a}_{\bf j}\rangle.
$$
Because
$v_3\psi_{j_z}^*\psi_{j_z-1}=i{|v_3|\over\sqrt{1+\om_z^2t^2}}\psi^0_{j_z}(\tilde
z)\psi^0_{j_z-1}(\tilde z)=-v_3^*\psi_{j_z}\psi_{j_z-1}^*$, the
second term cancels out. Then, without any approximation, $$
\langle \hat I({\bf r},t)\rangle={v_2(t)\over\prod\limits_\alpha
\sqrt{1+\om_\alpha^2t^2}}\;\rho_{eq}({\bf\tilde
r})=v_2(t)\rho({\bf r},t)$$ The flux is proportional to the
density of a cloud at thermal equilibrium with rescaled
coordinates. This means that the mean flux of an expanding
non-interacting cloud is proportional to the atomic density
without any approximation. This results holds with and without
gravity taken into account.

\subsubsection{Second order correlation}\label{2nd_order}

Here we calculate the correlation functions. A discussion is given in the next section.
The snap-shot correlation function is
$$G^{(2)}_{im.}({\bf r},t;{\bf r'},t)=\sum\limits_ {\bf j,k,l,n}
       \psi_{\bf j}^*\psi_{\bf k} \times{\psi'_{\bf l}}^*\psi'_{\bf n}~\langle
        \hat{a}_{\bf j}^\dag \hat{a}_{\bf k}\hat{a}_{\bf l}^\dag \hat{a}_{\bf n}\rangle.$$

\noindent Using Eq.(\ref{evol_temps}), one finds, without any approximation
(except the neglect of the shot-noise term):
$$G^{(2)}_{im}({\bf r},t;{\bf r'},t)=
{1\over\prod\limits_\alpha(1+\om_\alpha^2t^2)}\left(\rho_{eq}({\bf\tilde
r})\rho_{eq}({\bf\tilde r'})+|G^{(1)}({\bf\tilde r},{\bf\tilde
r'})|^2-\rho_0({\bf\tilde r})\rho_0({\bf\tilde r'})\right).
$$

\noindent As in the case of the mean density, the snap-shot
correlation function has the same form as in the trap except for
an anisotropic scale factor.

The calculation of $G^{(2)}_{fl.}$ is similar:
$$ \langle \hat I({\bf r},t)\hat I({\bf r'},t')\rangle=-\left({\hbar\over
2m}\right)^2
    \sum\limits_ {\bf j,k,l,n}
       [\psi_{\bf j}^*(\partial_z\psi_{\bf k})-(\partial_z\psi_{\bf j}^*)\psi_{\bf k}]
       \times[{\psi'_{\bf l}}^*(\partial_z\psi'_{\bf n})-(\partial_z{\psi'_{\bf l}}^*)\psi'_{\bf n}]~\langle
        \hat{a}_{\bf j}^\dag \hat{a}_{\bf k}\hat{a}_{\bf l}^\dag \hat{a}_{\bf n}\rangle
$$

\noindent Two major differences appear compared to the mean flux
calculation: the terms in $v_3$ and the phase factor
$\delta_\alpha +3\pi/2$ in Eq.(\ref{evol_temps}) do not cancel.
This makes the exact calculation very tedious. It is postponed to
the appendix.

Experiments are usually performed in situations satisfying two
conditions: (1) the width of the cloud after expansion is much
larger than that of the trapped cloud, and (2) the mean velocity
acquired during free fall is much larger than the velocity spread
of the trapped cloud. The first condition means that $\om_\alpha
t\gg 1$ and the second one that $gt\gg \sqrt{k_BT/m}$. The latter
condition also means that the mean arrival time,
$t_0=\sqrt{2H/g}$, is much larger than the time width
$\sqrt{k_BT/m g^2}$ of the expanding cloud. With these
approximations the scale factors become quite simple. $\tilde
x\sim\frac{x}{\om_x t_0},\tilde y\sim\frac{y}{\om_y t_0}$ and
$\tilde z\sim\frac{H-{1\over 2}gt^2}{\om_z t_0}\sim
\frac{g(t_0-t)}{\om_z}$. In particular, the coordinate $\tilde z$
is proportional to the arrival time $t$. This means that in
experiments that measure arrival times, the results have the same
form when expressed as a function of vertical position.

In the correlation function of the flux, the above approximations
also lead to $v_2\approx \sqrt{2gH}$ and
$|\sqrt{j_z}\;v_3/v_2|\approx
\sqrt{{k_BT\over\hbar\om_z}}{\si_z\over\sqrt{2}H}={s_z\over\sqrt{2}H}
$ where $s_z$ is the width of the cloud inside the trap and where
the typical value of the occupied trap level, $j_z$, is $\sim
{k_BT\over\hbar\om_z}$. The term containing $v_3$ is then very
small compared to the one proportional $v_2$. In
Ref.\cite{Schellekens} for instance the above ratio is $\sim
10^{-5}$. We will neglect terms containing $v_3$ in the following.
The phase factors $\delta_\alpha$ in Eq.(\ref{evol_temps}) are
also very small since $\om_\alpha t\gg 1$
and can be neglected (see appendix \ref{tt'}).\\

Under all these approximations, one finds
$$
G^{(2)}_{fl.}({\bf r},t;{\bf
r'},t')={v_2v'_2\over\prod\limits_\alpha\sqrt{(1+\om_\alpha^2t^2)(1+\om_\alpha^2t'^2)}}\left(\rho_{eq}({\bf\tilde
r})\rho_{eq}({\bf\tilde r'})+|G^{(1)}({\bf\tilde r},{\bf\tilde
r'})|^2-\rho_0({\bf\tilde r})\rho_0({\bf\tilde r'})\right)
$$
We again find the same correlation function as in the trap,
rescaled by
a  slightly different factor compared to
$G^{(2)}_{im.}$. This factor simply reflects the expansion of the
cloud between the times $t$ and $t'$.

The scaling laws for the harmonic potential result in a very
simple expression for the correlation lengths at the detector:

 \beq
l_\alpha^{(d)}={l^{(t)}} \times\sqrt{1+(\om_\alpha
t)^2}\label{xcoh}.\eeq

\noindent Where $l_\alpha^{(d)}$ is the correlation length along
the $\alpha$ direction at the detector and  $l_\alpha^{(t)}$ is
the correlation length in the trap. If the gas is far from
degeneracy $l^{(t)}={\lambda\over\sqrt{2\pi}}$, and we recover the
result of Eq. \ref{p coh length}. Close to degeneracy the
correlation length is position dependent. In the case of a pulse
of atoms as in Ref. \cite{Schellekens}, this formula applies along
all three space axes. In addition, when making a flux measurement,
one often expresses the longitudinal correlation length as a
correlation time. For a pulse of atoms from a harmonic trap,
with a mean velocity $v$ at the detector, the
correlation time is:

\beq t^{(coh)}={l_z^{(d)}\over v} ={l^{(t)}}
\times\frac{\om_z}{g}.\label{tcoh}\eeq
It is independent of the propagation time as long as $\om_zt\gg
1$.

\begin{figure}
\begin{center}
\includegraphics[scale=0.7]{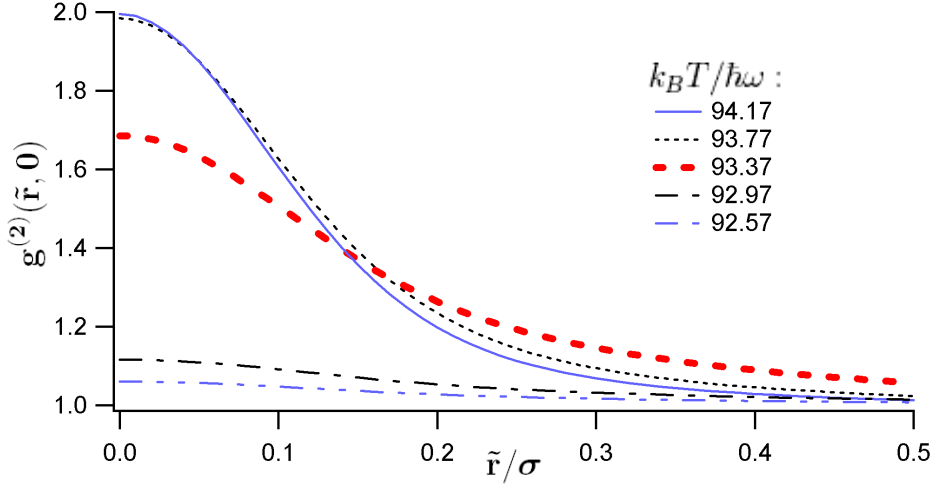}
\end{center}
\caption{(Color online) Two-body normalized correlation function
at the trap center, $g^{(2)}(\tilde r,0)$ for $10^6$ atoms as
function of the position $\tilde r=r/\om t$ for various
temperatures around transition temperature. The horizontal axis is
labelled in units of the size of the harmonic oscillator wave
function $\si$. The thick dashed line corresponds to the
transition temperature $T^*$ defined in Ref.\cite{Hoppeler} and is
$93.37~\hbar \omega/k_B$ for $10^6$ atoms. The temperature step is
$0.4~\hbar\omega/k_B$. The thermal de Broglie wavelength is $\sim
0.26~\si$. The effect of the ground state population is clearly
visible in the reduction of $g^{(2)}(0,0)$, and in the rapid
flattening out of the correlation function slightly below $T^*$.}
\label{g2}
\end{figure}

These calculations are illustrated in the following figures. For
simplicity we have used an isotropic trapping potential. As
pointed out above, the normalized second-order correlation
functions $g_{im.}^{(2)}$ and $g_{fl.}^{(2)}$ are virtually
identical with typical parameters (see appendix \ref{explicit
partII}) and we will use the shorter notation $g^{(2)}$. In Fig.
\ref{g2} we show the normalized correlation function
$g^{(2)}(\tilde r,0)$ as a function of $\tilde r\sim r/\om t$ for
various temperatures in the vicinity the Bose-Einstein phase
transition $T^*$. We use the saturation of the excited state
population to define $T^*$ \cite{Hoppeler}. This is the
correlation function \emph{at the center of the cloud}. One sees
that at $T=T^*$ (the thick dashed line in the figure), the
correlation function at zero distance is already significantly
diminished compared to higher temperatures. The correlation
length, on the other hand, is larger than $\lambda\om
t/\sqrt{2\pi}$. Also, one sees that the correlation function is
almost flat for temperatures a few percent below $T^*$.

\begin{figure}
\begin{center}
\includegraphics[scale=0.7]{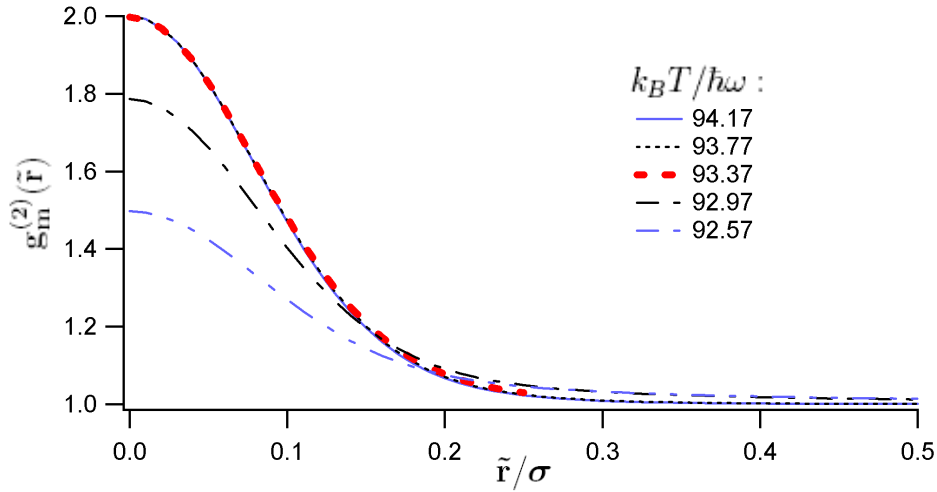}
\end{center}
\caption{(Color online) Two-body normalized correlation function
$g_m^{(2)}(\tilde r)$ for $10^6$ atoms as a function of $\tilde
r$. This function is an average of the two-body correlation
function over the cloud. The conditions are the same conditions as
for Fig.\ref{g2}. Unlike Fig.\ref{g2}, the shape is always almost
Gaussian and converges more slowly to a flat correlation for low
temperatures. This is because only a small region around $\tilde
r=0$ is fully sensitive to the quantum atomic distribution. }
\label{g2m}
\end{figure}

In many experiments of course, one does not measure the local
correlation function, but the correlation function averaged over
all points in the sample \cite{Schellekens}. The effect of this
averaging is shown in Fig. \ref{g2m}. We plot $g_m^{(2)}(\tilde
r)={\int d{\bf R}\; G^{(2)}({\bf R}+\tilde r {\bf e},{\bf
R})\over\int d{\bf R}\;G^{(1)}({\bf R}+\tilde r {\bf e},{\bf
R}+\tilde r {\bf e})G^{(1)}({\bf R},{\bf R})}$ where the vector
${\bf e}$ is a unit vector in some direction. One sees that the
amplitude of the correlation function decreases more slowly, and
that after averaging, the correlation length hardly varies as one
passes $T^*$.

\begin{figure}
\begin{center}
\includegraphics[scale=0.7]{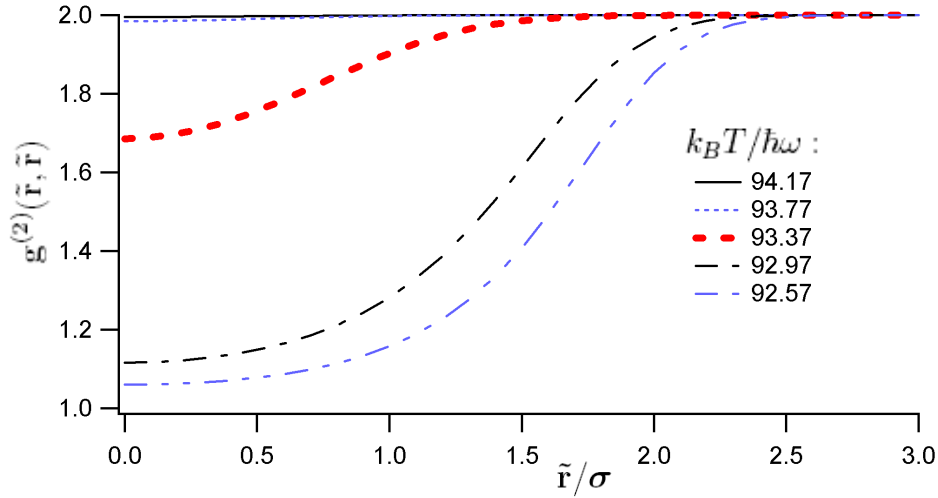}
\end{center}
\caption{(Color online) Two-body normalized correlation function
$g^{(2)}(\tilde r,\tilde r)$ for $10^6$ atoms as function of
$\tilde r$. The conditions are the same as for Fig.\ref{g2}. Even
for $T<T^*$ the correlation goes to 2 far from the center. This is
due to the finite spatial extent of the condensate. It can also be
understood in terms of the chemical potential $\mu(\tilde r)$
which, in a local density approximation, decreases as $\tilde r$
increases and thus the correlation is equivalent to that of a
hotter cloud.} \label{g2r}
\end{figure}

To illustrate how local the effects which distinguish Figs.
\ref{g2} and \ref{g2m} are, we also plot in Fig. \ref{g2r} the
value of $g^{(2)}(\tilde r,\tilde r)$, the zero distance
correlation function as a function of $\tilde r$ in the vicinity
of the cloud center. One sees that even below $T^*$, the
correlator is close to 2 at a rescaled distance of a few
times the harmonic oscillator length scale. We can simply
interpret this effect by observing that at $\tilde r$ the
effective chemical potential is $\mu - V(\tilde r)$. Away from the
center, the effective chemical potential is small and this part of
the cloud can be described as a Boltzmann cloud.

Before interpreting these results further, we recall some of our
assumptions and their possible violation. First, we obtain
Eq.(\ref{xcoh}) if we make a semi classical approximation assuming
that $k_BT$ greatly exceeds the energy spacing in the trap in each
dimension of space. In an anisotropic trap, this condition can be
violated in one or two dimensions and then correlation length
along these directions will be larger and can become infinite for
a small enough temperature. Second, we have assumed a
non-interacting gas throughout.

Repulsive interactions inflate the trapped cloud, and thus reduce
the length $l^{(d)}$ at the detector. We expect this to be the
main effect for atomic clouds above the Bose-Einstein transition
threshold, where the effects of atomic interactions are typically
small. The reduction is typically a few percent. Even slightly
below $T^*$, the condensate density is quite high, expelling the
thermal atoms from the center of the trap. The effects of
interactions inside the trap and during the cloud's expansion
cannot be neglected. Taking them into account is then complex and
beyond the scope of this paper.

\section{Physical interpretations}\label{sec interpret}

The main result of this paper is that in an experiment which
averages over a detector in the sense of Fig. \ref{g2m}, even at
$T=T^*$, the correlation lengths at the detector are well
approximated by:

$$l_\alpha^{(d)}={l^{(t)}} \times\om_\alpha t$$

\noindent The correlation length increases linearly with the time
of flight. A simple way to understand this result is to consider
the analogy with optical speckle. Increasing the time of flight
corresponds to increasing the propagation distance to the
observation plane in the optical analog. The speckle size, {\it
i.e.} the correlation length, obviously increases linearly with
the propagation distance. Another way to understand the time
dependence is to remark that after release, the atomic cloud is
free and the phase space density should be constant. Since the
density decreases with time as $\prod\limits_\alpha (\om_\alpha
t)$ and the spread of the velocity distribution is constant, the
correlation volume must increase by the same factor \cite{Miller}.

Yet another way to look at the correlation length is to observe
that, far from degeneracy, the correlation length inside the trap
is the thermal de Broglie wavelength, that is,
${\lambda\over\sqrt{2\pi}}= \hbar/\Delta p$ where $\Delta p=
m\Delta v$ is the momentum width of the cloud. By analogy, after
expansion, the correlation length is $\hbar/(\Delta p)_{loc}$,
where $(\Delta p)_{loc}$ is the ``local" width of the momentum
distribution. As the pulse of atoms propagates, fast and slow
atoms separate, so that at a given point in space the width in
momentum is reduced by a factor ${s_\alpha}\over{\Delta v t}$.

For a continuous beam, the formula (\ref{xcoh}) only applies in
the transverse directions. In the longitudinal direction, an
argument in terms of a local thermal de Broglie wavelength can be
used to find the coherence length or time. If the atoms travel at
velocity $v$ without acceleration, the momentum spread and
correlation length remain constant. Defining the energy width of
the beam as $\Delta E = m v \Delta v$, one finds a correlation
time $\lambda/ v =\hbar/\Delta E$ \cite{Iannuzzi}. In the presence
of an acceleration such as gravity, the momentum spread of the
beam decreases (the energy spread at any point $\Delta E$ is
constant), which increases the correlation length. The correlation
time, however, remains $\hbar/\Delta E$ \cite{Yasuda96}.

The result that the coherence length of a cloud of atoms can vary
with the distance of propagation, is in apparent contradiction
with the results of Refs. \cite{Kaiser83,Klein83}. Those papers
give convincing reasons, both experimental and theoretical, for
why the dispersion associated with the propagation of massive
particles should \emph{not} result in an increase of the coherence
length. The contradiction is resolved by noting that the
Mach-Zender interferometer considered in that work is sensitive to
the function $f({\bf r},t)=\int d{\bf R}\; G^{(1)}({\bf R},t;{\bf
R+r},t)$. If the Hamiltonian commutes with the momentum operator,
{\it i.e.} if plane waves are stationary states, one can easily
demonstrate that the function $f$ and hence its width are
independent of the time $t$. The experiments we analyze are
sensitive to the {\it modulus of $G^{(1)}$} whose width will
always increase with time. Thus the coherence length can depend on
the interferometer as well as the source.

The role of the acceleration of gravity in these experiments is
minor. It governs the propagation time and the speed of the
particles when they reach the detector. In a pulsed beam, gravity
has no effect on the correlation length, although it does affect
the correlation time. It also renders the rescaling of the $z$
coordinate linear for large times so that the correlation function
in position $z$ and time have the same form. Without gravity
(cancellation with a magnetic field gradient for example), a pulse
of atoms would take longer to reach the detector, thereby giving
the correlation length more time to dilate, and in addition they
would hit the detector at a lower velocity. The correlation time
would then increase with time and its order of magnitude would be
$\frac{\lambda\om t_0}{v_T}=\frac{\hbar\om}{k_BT} t_0$ where
$v_T=\sqrt{\frac{k_BT}{m}}$ is the thermal velocity and
$t_0=v_T/H$ is the time of flight to the detector.

\section{Effect of Finite Detector Resolution}\label{sec exp}

In the preceding sections, the detector was considered ideal, {\it i.e.} with arbitrarily good spatial and
temporal resolution. Here we will consider a model of a more realistic detector,
in which we suppose that the spatial resolution  in the $x-y$ plane is Gaussian.
This is often the case due to smearing in pixels
\cite{Bloch05, Esteve05} and is also approximately true in Ref. \cite{Schellekens}.
To simplify the discussion we will restrict our analysis to the case $T\gg T^*$
and use a Maxwell-Boltzmann distribution rather than Bose-Einstein
distribution. In this case, each direction of space is independent
and we will only consider one direction at a time in the following.

There are three different scales in the problem: the size of the
cloud at the detector $s(t)\approx \sqrt{\frac{k_BT}{m}}\,t$, the
correlation length at the detector $l^{(d)}$ and the r.m.s. width
of the detector resolution function $d$. The definition of the
resolution function is that for a density $\rho(x)=A e^{-{x^2\over
2 s(t)^2}}$, the observed density is given by a convolution:
$$\rho_{obs}(x)=\int dx_0 \rho(x_0){e^{-{1\over 2}({x-x_0\over
d})^2}\over \sqrt{2\pi}d} ={A\over\sqrt{1+ d^2/s(t)^2}}
e^{-{x^2\over 2[s(t)^2+ d^2]}}.$$

\noindent Similarly if $G^{(1)}(x,x')=A e^{i\phi} e^{-{(x+x')^2\over 2 (2
s)^2}}e^{-{(x-x')^2\over 2 (l^{(d)})^2}}$ is the first order
correlation function and $G_{obs}^{(1)}(x,x')$ the observed one,
we have

\begin{eqnarray}
|G_{obs}^{(1)}(x,x')|^2&=&\int dx_0 dx'_0
|G^{(1)}(x_0,x'_0)|^2\;{e^{-{1\over 2}({x-x_0\over d})^2}\over
\sqrt{2\pi}d}{e^{-{1\over 2}({x'-x'_0\over d})^2}\over
\sqrt{2\pi}d}\\
&=&{|A|^2 \over \sqrt{(1+d^2/s^2(t))(1+4
d^2/(l^{(d)})^2)}}e^{-{(x+x')^2\over 4[s(t)^2+
d^2]}}e^{-{(x-x')^2\over (l^{(d)})^2+4 d^2}}
\end{eqnarray}

Consequently, with $\alpha=x,y$ and $z$:

\bite

\item The amplitude of the normalized correlation function becomes

$g^{(2)}_{obs}({\bf 0,0})=\left(\frac{G_{obs}^{(1)}({\bf
0,0})}{\rho_{obs}({\bf 0})}\right)^2=
1+\prod\limits_\alpha\sqrt{1+d_\alpha^2/s^2_\alpha(t)\over
1+4d_\alpha^2/(l^{(d)}_\alpha)^2}$.

\item The observed widths of the cloud are $s_\alpha(t)\ra\sqrt{s_\alpha^2(t)+d_\alpha^2}$.

\item The observed correlation lengths are $l^{(d)}_\alpha\ra
\sqrt{(l^{(d)}_\alpha)^2+(2d_\alpha)^2}$. The factor $2$ can be
understood as
 $\sqrt{2}\times\sqrt{2}$ where the first term comes from
the fact that $d_\alpha$ is defined for one particle and not for a
pair of particles and the second one comes from the fact
that the correlation length is not defined as an r.m.s. width.

\eite

In the experiment of Ref. \cite{Schellekens} the
trapped cloud had a cigar shape. At the detector the cloud was
spherical but the correlation volume was anisotropic with
$l^{(d)}_x\ll d \approx  \l^{(d)}_y /4$. In the third (vertical) direction, the resolution width was
much smaller than any other length scale. The observed contrast of the correlation function was therefore approximately,
$ {l^{(d)}_x \over{2 d}}$.

\section{Conclusion}

The most important conclusion of this paper is that the expansion
of a non-interacting cloud from a harmonic trap in thermal
equilibrium, admits a rather simple, analytical treatment of the
time variation of the density and the correlation functions. In
such a pulse of atoms, correlation lengths scale in the same way
as the size of the density profile. The agreement with experiment
indicates that the neglect of interactions is a good approximation
above the BEC transition temperature. An important next step
however, is to examine interaction effects so that the next
generation of experiments, which will be more precise and better
resolved, can be fully interpreted.

\section*{Acknowledgments}

The Atom Optics group of LCFIO is member of the Institut
Francilien de Recherche sur les Atomes Froids (IFRAF) and of the
F\'ed\'eration LUMAT of the CNRS (FR2764). This work is supported
by the PESSOA program 07988NJ, by the Atom Chips network
MCRTN-CT-2003-505032, and the ANR under contract 05-NANO-008-01.

\section{Appendix}

\subsection{Explicit expression of the flux correlation function}
We found in section \ref{sec expansion}, the following expression
for the flux operator: $$ \hat I({\bf r},t)=
    \sum\limits_ {\bf j,k}
       \left[v_2\psi_{\bf j}^*\psi_{\bf k}-{1\over 2}\left(v_3\sqrt{k}\psi_{\bf j}^*\psi_{\bf k-1_z}
       +v_3^*\sqrt{j}\psi_{\bf j-1_z}^*\psi_{\bf k}  \right)\right]
       ~\hat{a}_{\bf j}^\dag \hat{a}_{\bf k}$$
where ${\bf j-1_z}$ is the vector $(j_x,j_y,j_z-1)$ and where we
write $\psi=\psi({\bf r},t)$.\\

The second order correlation function for the flux is then,

\btabul{l} $ \lan\hat I({\bf r},t)\hat I({\bf r'},t')\ran=$\\
$\sum\limits_ {\bf j,k,l,n}\left[v_2\psi_{\bf j}^*\psi_{\bf
k}-{1\over
2}\left(v_3\sqrt{k_z}\psi_{\bf j}^*\psi_{\bf k-1_z} +v_3^*\sqrt{j_z}\psi_{\bf j-1_z}^*\psi_{\bf k}  \right)\right]$\\
$\times\left[v'_2{\psi'}_{\bf l}^*\psi'_{\bf n}-{1\over
2}\left(v'_3\sqrt{n_z}{\psi'}_{\bf l}^*\psi'_{\bf n-1_z}+
{v'}_3^*\sqrt{l_z}{\psi'}_{\bf l-1_z}^*\psi'_{\bf n}
\right)\right] ~\lan\hat{a}_{\bf j}^\dag \hat{a}_{\bf
k}\hat{a}_{\bf l}^\dag \hat{a}_{\bf n}\ran$
\etabul\\

\noindent Neglecting the shot-noise and ground-state contributions, this
leads to

$\lan \hat I({\bf r},t)\hat I({\bf r'},t')\rangle=\langle \hat
I({\bf r},t)\ran\lan\hat I({\bf r'},t')\rangle + {\textrm Re}(A)$

\noindent with

\btabul{l} $A=\sum\limits_{\bf j,l} [v_2v'_2\, \psi_{\bf j}^*
\psi'_{\bf j} \psi_{\bf l} {\psi'}^*_{\bf l} +{1\over 2}v_3v'_3\,
\sqrt{j_zl_z}\,\psi_{\bf j}^* \psi'_{\bf j-1_z} \psi_{\bf l-1_z}
{\psi'}^*_{\bf l} +{1\over 2}v_3{v'}^*_3\, l_z\,\psi_{\bf j}^*
\psi'_{\bf j} \psi_{\bf l-1_z}
{\psi'}^*_{\bf l-1_z}$\\
$-v_2v'_3\, \sqrt{j_z}\,\psi_{\bf j}^* \psi'_{\bf j-1_z} \psi_{\bf
l} {\psi'}^*_{\bf l} -v'_2v_3\, \sqrt{l_z}\,\psi_{\bf j}^*
\psi'_{\bf j} \psi_{\bf l-1_z} {\psi'}^*_{\bf l}]\lan\hat{a}_{\bf
j}^\dag \hat{a}_{\bf j}\ran\lan\hat{a}_{\bf l}^\dag \hat{a}_{\bf
l}\ran$ \etabul\\

\noindent We write $A=\sum\limits_{i=1}^5T_i$ where the $T_i$ terms can be
recast, using $\tan\delta_\alpha=1/\om_\alpha t$,
$\tan\delta'_\alpha=1/\om_\alpha t'$,
$\Delta_\alpha=\delta'_\alpha-\delta_\alpha$, $\sum\limits_\alpha
j_\alpha(\delta'_\alpha-\delta_\alpha)={\bf j.\Delta}$,
$\psi^0_{\bf l}=\psi^0_{\bf l}({\bf\tilde r})$ and ${\psi'}^0_{\bf
l}=\psi^0_{\bf l}({\bf\tilde r'})$.

\bite

\item
$T_1=v_2v'_2\sum\limits_{\bf j,l} \psi_{\bf j}^* \psi'_{\bf j}
\psi_{\bf l} {\psi'}^*_{\bf l} \lan\hat{a}_{\bf j}^\dag
\hat{a}_{\bf j}\ran\lan\hat{a}_{\bf l}^\dag \hat{a}_{\bf l}\ran$

\btabul{l}
$=\frac{v_2v'_2}{\prod\limits_\alpha\sqrt{(1+\om_\alpha^2t^2)(1+\om_\alpha^2t'^2)}}\sum\limits_{\bf
j,l} \psi^0_{\bf j} {\psi'}^0_{\bf j} \psi^0_{\bf l}
{\psi'}^0_{\bf l}e^{i\sum\limits_\alpha
(j_\alpha-l_\alpha)(\delta'_\alpha-\delta_\alpha)}\lan\hat{a}_{\bf
j}^\dag \hat{a}_{\bf j}\ran\lan\hat{a}_{\bf l}^\dag \hat{a}_{\bf
l}\ran$\\
$=\frac{v_2v'_2}{\prod\limits_\alpha\sqrt{(1+\om_\alpha^2t^2)(1+\om_\alpha^2t'^2)}}\left|\sum\limits_{\bf
j} \psi^0_{\bf j} {\psi'}^0_{\bf j}\, e^{i{\bf
j.\Delta}}\lan\hat{a}_{\bf
j}^\dag \hat{a}_{\bf j}\ran\right|^2$\\

 \etabul

\item
$T_2={1\over 2}v_3v'_3\sum\limits_{\bf j,l}
\sqrt{j_zl_z}\,\psi_{\bf j}^* \psi'_{\bf j-1_z} \psi_{\bf l-1_z}
{\psi'}^*_{\bf l} \lan\hat{a}_{\bf j}^\dag \hat{a}_{\bf
j}\ran\lan\hat{a}_{\bf l}^\dag \hat{a}_{\bf l}\ran$

\btabul{l}
$=-\frac{1}{2}\frac{|v_3v'_3|}{\prod\limits_\alpha\sqrt{(1+\om_\alpha^2t^2)(1+\om_\alpha^2t'^2)}}\left(\sum\limits_{\bf
j} \sqrt{j_z}\psi^0_{\bf j} {\psi'}^0_{\bf j-1_z}e^{i{\bf
j.\Delta}} \lan\hat{a}_{\bf j}^\dag \hat{a}_{\bf
j}\ran\right)\left(\sum\limits_{\bf l}\sqrt{l_z}\psi^0_{\bf l-1_z}
{\psi'}^0_{\bf l}e^{-i{\bf l.\Delta}}\lan\hat{a}_{\bf l}^\dag
\hat{a}_{\bf l}\ran\right)$\etabul

\item
$T_3={1\over 2}v_3{v'}^*_3\,\sum\limits_{\bf j,l} l_z\,\psi_{\bf
j}^* \psi'_{\bf j} \psi_{\bf l-1_z} {\psi'}^*_{\bf l-1_z}
\lan\hat{a}_{\bf j}^\dag \hat{a}_{\bf j}\ran\lan\hat{a}_{\bf
l}^\dag \hat{a}_{\bf l}\ran$

\btabul{l}
$=\frac{1}{2}\frac{|v_3v'_3|}{\prod\limits_\alpha\sqrt{(1+\om_\alpha^2t^2)(1+\om_\alpha^2t'^2)}}\left(\sum\limits_{\bf
j} \psi^0_{\bf j} {\psi'}^0_{\bf j}e^{i{\bf j.\Delta}}
\lan\hat{a}_{\bf j}^\dag \hat{a}_{\bf
j}\ran\right)\left(\sum\limits_{\bf l}l_z\psi^0_{\bf l-1_z}
{\psi'}^0_{\bf l-1_z}e^{-i{\bf l.\Delta}}\lan\hat{a}_{\bf l}^\dag
\hat{a}_{\bf l}\ran\right)$\etabul

\item
$T_4=-v_2v'_3\, \sum\limits_{\bf j,l}\sqrt{j_z}\,\psi_{\bf j}^*
\psi'_{\bf j-1_z} \psi_{\bf l} {\psi'}^*_{\bf l}\lan\hat{a}_{\bf
j}^\dag \hat{a}_{\bf j}\ran\lan\hat{a}_{\bf l}^\dag \hat{a}_{\bf
l}\ran$

\btabul{l}
$=-i\frac{v_2|v'_3|}{\prod\limits_\alpha\sqrt{(1+\om_\alpha^2t^2)(1+\om_\alpha^2t'^2)}}\left(\sum\limits_{\bf
j} \sqrt{j_z}\psi^0_{\bf j} {\psi'}^0_{\bf j-1_z}e^{i{\bf
j.\Delta}} \lan\hat{a}_{\bf j}^\dag \hat{a}_{\bf
j}\ran\right)\left(\sum\limits_{\bf l}\psi^0_{\bf l}
{\psi'}^0_{\bf l}e^{-i{\bf l.\Delta}}\lan\hat{a}_{\bf l}^\dag
\hat{a}_{\bf l}\ran\right)$\\
\etabul

\item
$T_5=-v'_2v_3\,\sum\limits_{\bf j,l} \sqrt{l_z}\,\psi_{\bf j}^*
\psi'_{\bf j} \psi_{\bf l-1_z} {\psi'}^*_{\bf l}\lan\hat{a}_{\bf
j}^\dag \hat{a}_{\bf j}\ran\lan\hat{a}_{\bf l}^\dag \hat{a}_{\bf
l}\ran$

\btabul{l}
$=-i\frac{v'_2|v_3|}{\prod\limits_\alpha\sqrt{(1+\om_\alpha^2t^2)(1+\om_\alpha^2t'^2)}}\left(\sum\limits_{\bf
j} \psi^0_{\bf j} {\psi'}^0_{\bf j}e^{i{\bf j.\Delta}}
\lan\hat{a}_{\bf j}^\dag \hat{a}_{\bf
j}\ran\right)\left(\sum\limits_{\bf l}\sqrt{l_z}\psi^0_{\bf l-1_z}
{\psi'}^0_{\bf l}e^{-i{\bf l.\Delta}}\lan\hat{a}_{\bf l}^\dag
\hat{a}_{\bf l}\ran\right)$\\
\etabul

\eite
The term $T_1$ is a real number which is not the case for $T_2,
T_3, T_4$ and $T_5$.

\subsection{Calculation for harmonic oscillator stationary states}
All the above terms can be calculated analytically. All the series
are identical in the direction $x$ and $y$. We are then left with
the calculation of three series in only one direction: \bite
\item $\sum\limits_{n=0}^\infty\sqrt{n}\psi^0_{n-1}(\tilde z)\psi^0_n(\tilde z')e^{-nu}$
\item $\sum\limits_{n=0}^\infty\sqrt{n}\psi^0_{n}(\tilde z)\psi^0_{n-1}(\tilde z')e^{-nu}$
\item $\sum\limits_{n=0}^\infty {n}\psi^0_{n-1}(\tilde z)\psi^0_{n-1}(\tilde z')e^{-nu}$

\eite
The function $g_u(\tilde z,\tilde z')=\sum\limits_{n=0}^\infty
\psi^0_n(\tilde z)\psi^0_n(\tilde z')e^{-nu}$ is known
\cite{landau,glauber} and its expression is $g_u(\tilde z,\tilde
z')={1\over
\si\sqrt{\pi(1-e^{-2u})}}\exp[-\tanh(\frac{u}{2})\left({\tilde
z+\tilde z'\over 2\si}\right)^2-\coth(\frac{u}{2})\left({\tilde
z-\tilde z'\over 2\si}\right)^2]$.

Using $\tilde z\psi_n^0(\tilde z)={\si\over\sqrt{2}}\lan \tilde
z|\hat a + \hat
a^\dag|\psi_n^0\ran={\si\over\sqrt{2}}[\sqrt{n}\psi_{n-1}^0(\tilde
z)+\sqrt{n+1}\psi_{n+1}^0(\tilde z)]$, one finds
 $$\tilde
z g_u(\tilde z,\tilde z')={\si\over\sqrt{2}}[\sum
\sqrt{n}\psi^0_{n-1}(\tilde z)\psi^0_n(\tilde z')e^{-nu}+e^u\sum
\sqrt{n}\psi^0_{n}(\tilde z)\psi^0_{n-1}(\tilde z')e^{-nu}].$$

\noindent It follows easily that \bite
\item $\sum\limits_{n=0}^\infty\sqrt{n}\psi^0_{n-1}(\tilde z)\psi^0_n(\tilde z')e^{-nu}={\sqrt{2}\over\si}{\tilde z-e^u\tilde z'\over 1-e^{2u}}
g_u(\tilde z,\tilde z')$
\item $\sum\limits_{n=0}^\infty\sqrt{n}\psi^0_{n}(\tilde z)\psi^0_{n-1}(\tilde z')e^{-nu}={\sqrt{2}\over\si}
{\tilde z'-e^u\tilde z\over 1-e^{2u}}g_u(\tilde z,\tilde z')$
\eite

\noindent Moreover, $\sum\limits_{n=0}^\infty {n}\psi^0_{n-1}(\tilde
z)\psi^0_{n-1}(\tilde z')e^{-nu}=e^{-u}[g_u(\tilde z,\tilde
z')-\partial_u g_u(\tilde z,\tilde z')]$. Then, \bite
\item $\sum\limits_{n=0}^\infty
{n}\psi^0_{n-1}(\tilde z)\psi^0_{n-1}(\tilde z')e^{-nu}=[{1\over
1-e^{-2u}}+{1\over 2}({\tilde z+\tilde z'\over
2\si\cosh{\frac{u}{2}}})^2- {1\over 2}({\tilde z-\tilde z'\over
2\si\sinh{\frac{u}{2}}})^2]e^{-u}g_u(\tilde z,\tilde z')$ \eite

\subsection{Explicit expression of the flux correlation function-Part
II}\label{explicit partII}

We define $G^{(1)}_B({\bf r},{\bf r'},{\bf u})=\sum\limits_{\bf
n}^\infty \psi^0_{\bf n}({\bf r})\psi^0_{\bf n}({\bf r'})e^{-{\bf
n u}}$. This function, the 3D equivalent of the function $g_u$, is
connected to the one-body correlation function by $G^{(1)}({\bf
r},{\bf r'})=\sum\limits_{l=1}^\infty e^{\beta l\tilde\mu}
G^{(1)}_B({\bf r},{\bf r'},l\tauv)$ with
$\tau_\alpha={\beta\hbar\om_\alpha}$.

Then,

\bite

\item $T_1=\frac{v_2v'_2}{\prod\limits_\alpha\sqrt{(1+\om_\alpha^2t^2)(1+\om_\alpha^2t'^2)}}
\left|\sum\limits_l e^{\beta l \tilde\mu}G^{(1)}_B({\bf\tilde
r},{\bf\tilde r'},l\tauv-i\bf\Delta)\right|^2$

\item $T_2=-\frac{1}{2}\frac{|v_3v'_3|}{\prod\limits_\alpha\sqrt{(1+\om_\alpha^2t^2)(1+\om_\alpha^2t'^2)}}
\left(\sum\limits_l e^{\beta l \tilde\mu}{\sqrt{2}\over\si}{\tilde
z-e^{l\tau_z-i\Delta_z}\tilde z'\over
1-e^{2(l\tau_z-i\Delta_z)}}\,G^{(1)}_B({\bf\tilde r},{\bf\tilde
r'},l\tauv-i\bf\Delta)\right)$

$\times\left(\sum\limits_k e^{\beta k
\tilde\mu}{\sqrt{2}\over\si}{\tilde z'-e^{k\tau_z+i\Delta_z}\tilde
z\over 1-e^{2(k\tau_z+i\Delta_z)}}\,G^{(1)}_B({\bf\tilde
r},{\bf\tilde r'},k\tauv+i\bf\Delta)\right)$

\item $T_3=\frac{1}{2}\frac{|v_3v'_3|}{\prod\limits_\alpha\sqrt{(1+\om_\alpha^2t^2)(1+\om_\alpha^2t'^2)}}\left(\sum\limits_l e^{\beta l \tilde\mu}\,G^{(1)}_B({\bf\tilde r},{\bf\tilde
r'},l\tauv-i\bf\Delta)\right)$

$\times\left(\sum\limits_k e^{\beta k \tilde\mu}[{1\over
1-e^{-2(k\tau_z+i\Delta_z)}}+{1\over 2}({\tilde z+\tilde z'\over
2\si\cosh{\frac{{k\tau_z+i\Delta_z}}{2}}})^2- {1\over 2}({\tilde
z-\tilde z'\over
2\si\sinh{\frac{{k\tau_z+i\Delta_z}}{2}}})^2]e^{-(k\tau_z+i\Delta_z)}\,G^{(1)}_B({\bf\tilde
r},{\bf\tilde r'},k\tauv+i\bf\Delta)\right)$

\item $T_4=-i\frac{v_2|v'_3|}{\prod\limits_\alpha\sqrt{(1+\om_\alpha^2t^2)(1+\om_\alpha^2t'^2)}}\left(\sum\limits_l e^{\beta l \tilde\mu}\,{\sqrt{2}\over\si}{\tilde
z-e^{l\tau_z-i\Delta_z}\tilde z'\over
1-e^{2(l\tau_z-i\Delta_z)}}\,G^{(1)}_B({\bf\tilde r},{\bf\tilde
r'},l\tauv-i\bf\Delta)\right)$

$\times\left(\sum\limits_k e^{\beta k
\tilde\mu}\,G^{(1)}_B({\bf\tilde r},{\bf\tilde
r'},k\tauv+i\bf\Delta)\right)$

\item $T_5=-i\frac{|v_3|v'_2}{\prod\limits_\alpha\sqrt{(1+\om_\alpha^2t^2)(1+\om_\alpha^2t'^2)}}\left(\sum\limits_l e^{\beta l \tilde\mu}\,G^{(1)}_B({\bf\tilde r},{\bf\tilde
r'},l\tauv-i\bf\Delta)\right)$

$\times\left(\sum\limits_k e^{\beta k
\tilde\mu}\,{\sqrt{2}\over\si}{\tilde
z'-e^{k\tau_z+i\Delta_z}\tilde z\over
1-e^{2(k\tau_z+i\Delta_z)}}\,G^{(1)}_B({\bf\tilde r},{\bf\tilde
r'},k\tauv+i\bf\Delta)\right)$ \eite

\noindent The dominant term is $T_1$ and is the one used in section
\ref{2nd_order}.

\subsection{Contribution of neglected terms in the correlation of the flux}

Here we evaluate the neglected the terms $T_2$ to
$T_5$ and the shot-noise contribution.
They will be evaluated in the case of clouds far above BEC
threshold. Under this assumption, all the functions are separable
in the variables $x,y$ and $t$ and the summation over the index
$l$ in the previous equations reduces to the single term $l=1$.

\subsubsection{Shot-noise contribution}

Using the above analysis one can show that the main term is still
proportional to  $v_2v'_2$. The additional term is then,
$$\frac{v_2v'_2}{\prod\limits_\alpha\sqrt{(1+\om_\alpha^2t^2)(1+\om_\alpha^2t'^2)}}e^{\beta\tilde\mu}
G^{(1)}_B({\bf\tilde r},{\bf\tilde
r'},\tauv-i{\bf\Delta})G^{(1)}_B({\bf\tilde r},{\bf\tilde
r'},i{\bf\Delta})$$

For $t=t'$, $\bf\Delta={\bf 0}$ and $G^{(1)}_B({\bf\tilde
r},{\bf\tilde r'},\bf 0)=\delta(\bf\tilde r-\tilde r')$. The
shot-noise term is then
$$\frac{v_2^2}{\prod\limits_\alpha(1+\om_\alpha^2t^2)}\rho_{eq}({\bf\tilde r})\delta(\bf\tilde r-\tilde r')$$

As expected, this term corresponds also to the one at equilibrium
with rescaled coordinates.

\subsubsection{$T_2-T_5$ contribution}\label{tt'}

We have $G^{(2)}_{fl.}({\bf r},t;{\bf r'},t')=\lan \hat I({\bf
r},t)\hat I({\bf r'},t')\rangle=\langle \hat I({\bf
r},t)\ran\lan\hat I({\bf r'},t')\rangle + {\textrm Re}(A)$ where
$A=\sum\limits_{i=1}^5 T_i$

\bite

\item Case $t=t'$.

\bite

\item $\bf\Delta=0$,

\item then $T_1=\frac{v_2v'_2}{\prod\limits_\alpha\sqrt{(1+\om_\alpha^2t^2)
(1+\om_\alpha^2t'^2)}}\left|G^{(1)}({\bf\tilde
r},{\bf\tilde r'})\right|^2$, $T_2$ and $T_3$ are real number and
${\textrm Re}(T_4)={\textrm Re}(T_5)=0$.

\item One finds, to leading orders,

$g^{(2)}(0,0,t;0,0,t)-2 \approx {1\over 8}\left({s_z\over
H}\right)^2(1-2{t-t_0\over t_0})(1-{\tau_z^2\over 6}) $ where
$s_z$ is the initial size of the cloud in the vertical direction
and $t_0=\sqrt{2H/g}$.

\item The deviation from 2 is extremely small in the experimental
conditions of \cite{Schellekens} ($\sim~10^{-11}$) but shows that
the bunching is strictly speaking not 2 at the center. This
behavior is expected for any flux correlation function of
dispersive waves \cite{flux-continu}.

\item The correlation lengths at the detector are not modified by the additional terms.

\eite

\item Case $t\neq t'$.

\bite

\item The correlation function can be written as

$g^{(2)}(0,0,t;0,0,t')=1 +\frac{|G^{(1)}_B({\bf\tilde
r},{\bf\tilde r'},\tauv+i\bf\Delta)|^2}{G^{(1)}_B({\bf\tilde
r},{\bf\tilde r},\tauv)G^{(1)}_B({\bf\tilde r'},{\bf\tilde
r'},\tauv)}[1+\eps]$.

\item where $\frac{|G^{(1)}_B({\bf\tilde
r},{\bf\tilde r'},\tauv+i\bf\Delta)|^2}{G^{(1)}_B({\bf\tilde
r},{\bf\tilde r},\tauv)G^{(1)}_B({\bf\tilde r'},{\bf\tilde
r'},\tauv)}\approx e^{-\left({t-t'\over
t^{(coh)}}\right)^2(1-{\tau_z^2\over 6})[1-({t+t'-2t_0\over
t_0})]}$ and

\item $\eps\approx {1\over 8}\left({w_z\over
H}\right)^2[1-({t+t'-2t_0\over t_0})](1-{\tau_z^2\over 6})-{3\over
2(\om_zt_0\tau_z)^2}\left({t-t'\over t_0}\right)^2(1+{\tau_z\over
3})$.

We have neglected terms in $\tau_z, (t-t_0)^3, (t'-t_0)^3,
(t-t_0)^2(t'-t_0),(t-t_0)(t'-t_0)^2$ and higher orders.

\item The value of $\eps$ is extremely small ($\sim 10^{-10}$) using Ref.\cite{Schellekens}. The deviation from $e^{-\left({t-t'\over
t^{(coh)}}\right)^2}$ is mainly due to the mean time $(t+t')/2$
contribution and changes the value of the correlation time in the
wings of the time-of-flight by $\sim 3~\%$. The effect of the
phase factor $\bf\Delta$ is negligible.

\eite

\eite

\end{document}